\documentclass[aps,prl,superscriptaddress,twocolumn,10pt]{revtex4-2}

\usepackage{verbatim}
\usepackage{amssymb}
\usepackage{cancel}
\usepackage{xcolor,graphicx}
\usepackage{amsmath}
\usepackage{amsbsy}
\usepackage{amsthm}
\usepackage{bbm}
\usepackage{epsfig}
\usepackage{float}
\usepackage{subfigure}
\usepackage{dcolumn}
\usepackage{epstopdf}
\usepackage{amscd}
\usepackage{amsfonts}
\usepackage{mathrsfs}
\usepackage{verbatim}
\usepackage[]{cases}
\usepackage{wasysym}
\usepackage[latin9]{inputenc}
\usepackage[T1]{fontenc}
\usepackage{mathtools}
\usepackage{todonotes}
\usepackage[normalem]{ulem}
\usepackage[colorlinks=true,linkcolor=blue,urlcolor=blue,citecolor=blue,pdfusetitle]{hyperref}
\usepackage{cleveref}
\usepackage{libertine}
\usepackage{braket}

\newcommand{\be}{\begin{equation}}
\newcommand{\e}[1]{\label{#1}\end{equation}}
\def\bea{\begin{eqnarray}}
\def\ea#1{\label{#1}\end{eqnarray}}
\def\rqn#1{(\ref{#1})}
\def\ee{\end{equation}}
\def\eea{\end{eqnarray}}
\def\bes#1{\begin{subequations}\label{#1}}
\def\ese{\end{subequations}}

\newcommand{\Tr}{\operatorname{Tr}}

\renewcommand{\ket}[1]{\arrowvert #1 \rangle} 
\renewcommand{\bra}[1]{\langle #1 \arrowvert} 

\newcommand{\ABS}[1]{\arrowvert #1 \arrowvert^2} 

\newcommand{\ud}{\mathrm{d}}
\newcommand{\un}[1]{\ensuremath{\mathrm{\,#1}}}

\newcommand{\blue}[1]{{\color{black}{#1}}}

\newcommand{\new}[1]{{\color{black}{#1}}}

\begin{document}

\title{Sensing microscopic noise events by frequent quantum measurements }

\author{Salvatore Virz\`i}
\affiliation{Istituto Nazionale di Ricerca Metrologica, Strada delle Cacce 91, 10135, Torino, Italy}
\author{Laura T. Knoll}
\affiliation{Istituto Nazionale di Ricerca Metrologica, Strada delle Cacce 91, 10135, Torino, Italy}
\affiliation{DEILAP-UNIDEF, CITEDEF-CONICET, J.B. de La Salle 4397, 1603 Villa Martelli, Buenos Aires, Argentina}
\author{Alessio Avella}
\author{Fabrizio Piacentini}
\affiliation{Istituto Nazionale di Ricerca Metrologica, Strada delle Cacce 91, 10135, Torino, Italy}
\author{Stefano Gherardini}
\affiliation{Istituto Nazionale di Ottica del Consiglio Nazionale delle Ricerche (CNR-INO), Area Science Park, Basovizza, I-34149 Trieste, Italy}
\author{Marco Gramegna}
\affiliation{Istituto Nazionale di Ricerca Metrologica, Strada delle Cacce 91, 10135, Torino, Italy}
\author{Gershon Kurizki}
\author{Abraham G. Kofman}
\affiliation{Department of Chemical and Biological Physics, Weizmann Institute of Science, Rehovot 7610001, Israel}
\author{Ivo Pietro Degiovanni}
\author{Marco Genovese}
\affiliation{Istituto Nazionale di Ricerca Metrologica, Strada delle Cacce 91, 10135, Torino, Italy}
\affiliation{INFN, sez. di Torino, via P. Giuria 1, 10125, Torino, Italy}
\author{Filippo Caruso}
\affiliation{Department of Physics and Astronomy and European Laboratory for Non-Linear Spectroscopy (LENS), University of Florence, via G. Sansone 1, 50019 Sesto Fiorentino, Italy}

\begin{abstract}
We propose and experimentally demonstrate a general method allowing us to unravel microscopic noise events that affect a continuous quantum variable.
Such unraveling is achieved by frequent measurements of a discrete variable coupled to the continuous one.
The experimental realization involves photons traversing a noisy channel.
There, their polarization, whose coupling to the photons' spatial wavepacket is subjected to stochastic noise, is frequently measured in the quantum Zeno regime.
The measurements not only preserve the polarization state, but also enable the recording of the full noise statistics from the spatially-resolved detection of the photons emerging from the channel.
This method proves the possibility of employing photons as quantum noise sensors and robust carriers of information.
\end{abstract}

\maketitle

The existing approaches to sensing by quantum probes \cite{ChernobrodJAP2005,TaylorNatPhys2008,MazeNature2008,BalasubramanianNature2008,MaiwaldNatPhys2009,BylanderNatPhys2011,KohlhaasPRX2015,DegenRMP2017,SchioppoNatPhot2017,ReiterNatComm2017,FreyNatComm2017,PezzeRMP2018,HernandezPRB2018,SungNatComm2019,McCormickNature2019,D1,D2,D3} commonly consider the noise \cite{PalmaPRSLSA1996,GardinerBook2000,BreuerBook2002,BenattiIJMP2005,ViolaPRL2010,CarusoRMP2014,ManisPRA2016,ParisPRA2017} affecting the probed observables as a nuisance to be suppressed \cite{KKbook}.
This can be done either  by active dynamical control of the probe \cite{22a,qco1,qco2,KKbook} or (if possible) by confining the sensing to a decoherence-free subspace of the probed object \cite{ZanardiPRL1997,LidarPRL1998,KwaitScience2000}.
Yet, noise sensing  by a quantum probe can be a source of valuable information on the underlying stochastic processes and their detrimental effect on quantum  coherence, which is a central obstacle to quantum information technologies \cite{DowlingPTRSA2003,OBrienNatPhot2009,PNASKur,StreltsovRMP2017,SafronovaRMP2018,WangNatPhot2020,AdamsReview2020,Geno2021,ReviewHorodecki}.
Existing noise sensing focuses on noise spectroscopy by qubit probes \cite{AlvarezPRL2011,YugePRL2011,Paz-SilvaPRL2014,q11,NorrisPRL2016,MuellerSciRep2016,SzankowskiJPCM2017,MuellerSciRep2018,KrzywdaNJP2019,DoNJP2019,SakuldeePRA2020,MuellerPLA2020}, whose dynamical control can enhance their ability to extract noise characteristics, such as the noise memory time \cite{q11}.
However, all such methods are sensitive only to statistical averages over many noise realizations, and are restricted by two major assumptions that severely limit their applicability \cite{KKbook}:  i) the noise is a stationary process (which is often untrue); ii) the probe and the bath, which is the source of noise, remain uncorrelated, and the bath state is unchanged by their interaction, i.e. the Born approximation holds.\\
Here we venture into the scarcely explored domain of {\em probing individual microscopic noise events} without the restrictions of noise stationarity and the Born approximation.
To this end, we resort to {\em frequent projective measurements} of the probe at intervals that are shorter than (or comparable to) those of noise events.
As demonstrated here both theoretically and experimentally, such measurements allow the unraveling of the full statistics of individual noise events, thus providing information impossible to obtain with existing noise sensing methods.\\
Frequent measurements have been employed to slow down the evolution of quantum systems coupled to baths, thus protecting them from relaxation or decoherence when the measurement rate conforms to the quantum Zeno effect (QZE) \cite{KKbook,MuellerSciRep2016,MisraJMP1977,ItanoPRA1990,4,QZEhome,FacchiPRL2002,QZEks,QZEmat,SchaferNatComm2014,SignolesNatPhys2014,GherardiniNJP2016}.
We have recently demonstrated \cite{QZE-AZE-prl2022} that polarization measurements of a photonic probe can disclose the sign of correlations between consecutive polarization fluctuations in a noisy medium (bath) that adheres to the Born approximation. 
Here we consider a far more general scenario, where the probe and the bath become entangled by each noise event (inducing decoherence on the system state), and the probe measurements keep changing the (continuous-variable) bath state.
An appropriate probe-measurement rate conforming to the QZE is shown to reveal the full distribution of random probe-bath couplings.
This hitherto unavailable information can be used to devise novel strategies of noise sensing and control, especially in communication channels.

\textit{Noise sensing by frequent probe measurements ---}
Consider a quantum system coupled to a bath via the interaction Hamiltonian
\be
H_{\rm I}=-\kappa(t)S \otimes B,
\e{1}
where $S$ and $B$ are the operators of the system (S) and the bath (B), respectively, $\kappa(t)$ is the time-dependent (here - stochastic) coupling strength, and we assume that the Hamiltonian of both S and B vanishes.\\
The combined \textit{supersystem} (S+B) evolves via the unitary operator $U(t)=e^{iG(t)S\otimes B}$, where $G(t)=\int_0^td\tau\,\kappa(\tau)$.
Let the initial state of the supersystem be $\ket{\Psi_{\rm in}}=\ket{\psi}\otimes\ket{f}$, where $\ket{\psi}$ and $\ket{f}$ are the initial states of S and B, respectively.
The probability that S will remain in its initial state is given by $p(t)=\Tr(\Pi_{\psi}\ket{\Psi(t)}\bra{\Psi(t)})$, where $\Pi_{\psi}=\ket{\psi}\bra{\psi}$ and $\ket{\Psi(t)}=U(t)\ket{\Psi_{\rm in}}$.
For sufficiently small $t$, we can expand the exponential in $U(t)$ up to second order, yielding 
\be
p(t)=1-G^2(t)\Delta S^2\,\overline{B^2},
\e{7}
where $\overline{B^2}=\braket{f|B^2|f}$, $\overline{S^n}=\braket{\psi|S^n|\psi}$ and $\Delta S^2=\overline{S^2}-\overline{S}^{\,2}$ is the variance of $S$ in the state $\ket{\psi}$.
Assume that the system S is measured in the state $\ket{\psi}$ at {\em random} instants $t_j$ in the time interval $[0,T]$ and, immediately after each $t_j\ (j=1,\dots,N;\ t_N=T)$, it is coupled to a new bath in the same state $\ket{f}$.
If the intervals $\tau_j=t_j-t_{j-1}$ (where $t_0=0$) are sufficiently short, then the total survival probability of the system state $\ket{\psi}$ is a product of expressions of the form of Eq. \rqn{7}, yielding
\be
p_{\rm tot}=e^{-J_N},\quad
J_N=\Delta S^2\,\overline{B^2}\,\sum_{j=1}^Ng_j^2,
\e{8}
where $g_j=\int_{t_{j-1}}^{t_j}dt\,\kappa(t)$.\\
For simplicity, we assume that the stochastic $\kappa(t)$ does not change sign, i.e. $\kappa(t)\ge0$.
Consider first the effect of unitary evolution, so that there is only one measurement at the end of the process, at $t_1=T$.
Then
\be
J_1=\Delta S^2\,\overline{B^2}\,G^2(T)=
\Delta S^2\,\overline{B^2}\,\left(\sum_{j=1}^Ng_j\right)^2,
\e{13}
where we used the equality $G(T)=\sum_{j=1}^Ng_j$.
Thus, $J_1$ [Eq.\ \rqn{13}] contains $N^2$ terms, which are roughly of the same order.
By comparison, for $N$ measurements during time $T$, $J_N$ in Eq.\ \rqn{8} contains $N$ terms, which are a subset of the terms in Eq.\ \rqn{13}, implying that $J_N/J_1\sim1/N$.\\
Hence, the slowdown of the system state decay under $N$ stochastic system-bath coupling events that are interrupted by $N$ projections on the initial state is similar to the well-known QZE scaling in the case of constant coupling \cite{KKbook}.\\
The crucial insight transpiring from Eqs.\ \rqn{8} and \rqn{13} is that the decay of the system state in the stochastic QZE regime \cite{GherardiniNJP2016} depends on the sum of the {\em squared random couplings} $g_j^2$, as opposed to its dependence on the sum of $g_j$ in $G(T)$ if the decay is measured at the end of the evolution during $T$.
Hence, {\em a novel, unfamiliar role of projective measurements in the QZE regime}, beyond protecting the system state from decoherence, is revealed by this analysis: their ability to provide {\em information on the distribution of random, microscopic  decoherence events} induced by the coupling of the system to the bath.

%
\textit{Continuous-variable noise unraveling by polarization probe measurements ---}
We will now show that these results allow unraveling noise events that affect the photon polarization qubit, considered as a probe system coupled to a bath realized by a transverse spatial degree of freedom (DoF) of the photon.
Namely, the photonic qubit propagates through a noisy channel in which decoherence occurs via coupling of the qubit with a spatial, continuous variable of the photon, acting as the bath.
The propagation is a sequence of steps, each associated with a noise event; in the $j$-th step, the coupling of the qubit with the continuous DoF is realized by the coupling of the polarization to the transverse position $x$ of the photon, $z$ being the propagation axis, with strength randomly changing at each step.\\
Specifically, the photonic qubit is dephased in the basis $\{\ket{H},\ket{V}\}$, $H$ and $V$ denoting the horizontal and vertical polarizations, respectively.
The decoherence manifests itself as a spatial mismatch between the $H$ and $V$ polarization components: $H$ polarization is (slightly) spatially shifted along the $x$ axis by the unitary operator $U_j=\exp(ig_j P_x\otimes\Pi_H)$, $P_x$ denoting the transverse momentum along the $x$ direction and $\Pi_H=\ket{H}\bra{H}$.\\
In the general system-bath notation, $S=\Pi_H$ and $B=P_x$.
The initial joint state of the qubit and the bath is
\be
\ket{\Psi_{\rm in}}=\ket{\psi_\theta}\otimes\ket{f_x},\quad
\ket{\psi_\theta}=\cos(\theta)\ket{H}+\sin(\theta)\ket{V},
\e{e6}
where $\ket{f_x}$ corresponds to the Gaussian wavepacket $f(x)=\braket{x|f_x}=\frac{1}{(2\pi\sigma^2)^{1/4}}\exp\left(-\frac{x^2}{4\sigma^2}\right)$.
For the states in Eq. \rqn{e6},
\be
\Delta S^2=\sin^2(\theta)\cos^2(\theta),\quad\overline{B^2}=\frac{1}{2\sigma^2}.
\e{18}

Since, in general, the bath state changes after each measurement, the decay parameter in \rqn{8} is now given by
\be
J_N=\sin^2(\theta)\cos^2(\theta)\,\sum_{j=1}^Ng_j^2\,\overline{B^2}_j.
\e{20}
Here $\overline{B^2}_j$ is calculated for the spatial DoF wavepacket evolving between the $(j-1)$th and $j$th measurements, whereas $\overline{B^2}_1=\overline{B^2}$.
According to Eq. \rqn{18}, $\overline{B^2}$ is inversely proportional to the square of the wavepacket width.
Given that $B^2=P_x^2$ applies the second derivative (with respect to $x$) to the wavepacket, $\overline{B^2}_j$ generally decreases with the wavepacket broadening.
In our case, the wavepacket evolution consists in its splitting into different sub-packets which recede from each other, resulting, as long as the distance between the sub-packets is less than their widths, in a broadening of the wavepacket.
Therefore, the quantities $\overline{B^2}_j$ in \rqn{20} are expected to decrease with $j$, causing a decrease of the decay parameter $J_n$, {\em in addition to the QZE}.

In our scenario, the quantum channel hosts $N$ noise events, each characterized by a coupling strength randomly chosen from a finite set of values, $g_j\in\{G_1,\dots,G_D\}$, with the probability $p_k$ of the $G_k$ noise event being independent of the events order.
In such a channel, the probability of observing $\{n_1,\dots,n_D\}$ events of the type $\{G_1,\dots,G_D\}$ is given by the multinomial distribution $M(n_1,\dots,n_D)=\frac{N!}{n_1!\dots n_D!}\prod_{k=1}^Dp_k^{n_k}$, where $\sum_{k=1}^Dn_k=N$ and $\sum_{k=1}^Dp_k=1$.

The corresponding (non-normalized) output state is:
\be
\ket{\Psi_{\rm out}}=\Pi_\theta U_N\dots
\Pi_\theta U_1\ket{\Psi_{\rm in}}=
\ket{\psi_\theta}\otimes\prod_{j=1}^NB_j\ket{f_x},
\e{e2}
where $B_j\equiv \cos^2(\theta)e^{ig_jP_x}+\sin^2(\theta)I_d$, $I_d$ being the identity operator in the Hilbert space of the spatial DoF.

Since the $B_j$ operators commute with each other, we can have $\prod_{j=1}^N B_j = \prod_{k=1}^D \left(\cos^{2}(\theta)\,e^{iG_k P_x} + \sin^{2}(\theta)\,\mathbb{I}_d\right)^{n_k}$,
highlighting how the spatial profile of $\ket{\Psi_{\mathrm{out}}}$ only depends on the
values of $G_k$ and their multiplicity, not on their ordering.
Hence, the probability of finding the photon in a specific position $x_0$ at the output of the quantum channel is
\begin{equation}
\mathcal{P}_N(x_0,\{n_k\}) =
\frac{\ABS{\braket{x_0}{\prod_{k=1}^D B_k^{n_k}}{f_x}}}{\int \ud x' \, \ABS{\braket{x'}{\prod_{k=1}^D B_k^{n_k}}{f_x}}}\,.
\end{equation}
Since we are interested in the estimation of the probability set $\{p_k\}$, we need to quantify the numbers $\{n_k\}$ of the decoherence events $\{G_k\}$.
This is achieved by minimizing, for any position $x$, the statistical square distance $\Delta \mathcal{P}(\{n_k\}_c)\equiv	\int\! \ud x \, \left[\mathcal{P}^{\mathrm{exp}}_N(x) - \mathcal{P}_N(x,\{n_k\}_c)\right]^{2}$, $\mathcal{P}^{\mathrm{exp}}_N(x)$ being the distribution experimentally measured at the output of the quantum channel and $\mathcal{P}_N(x,\{n_k\}_c)$ the theoretical one given by the set $\{n_k\}_c$, with the integer index $c \in \left[1,\binom{D + N - 1}{N}\right]$ labeling the theoretical distributions.
The reconstructed set $\{n_k\}_R$ of the multiplicities of decoherence events is estimated as
%
\begin{equation}\label{setN}
\{n_k\}_R = {\rm arg}\min_{c} \Delta \mathcal{P}(\{n_k\}_c),
\end{equation}
yielding the probabilities $\{p_k\}_R$, with $p_k = n_k/N$.
To reduce the uncertainty on the estimated $p_k$ values, one can repeat this procedure $L$ times, thus obtaining each $p_k$ as $p_k = \sum_{\ell=1}^{L}p_k^{(\ell)}/L$.

{\it The experiment --- }
\begin{figure}[t!]
	\centering
	\includegraphics[width=.95\columnwidth]{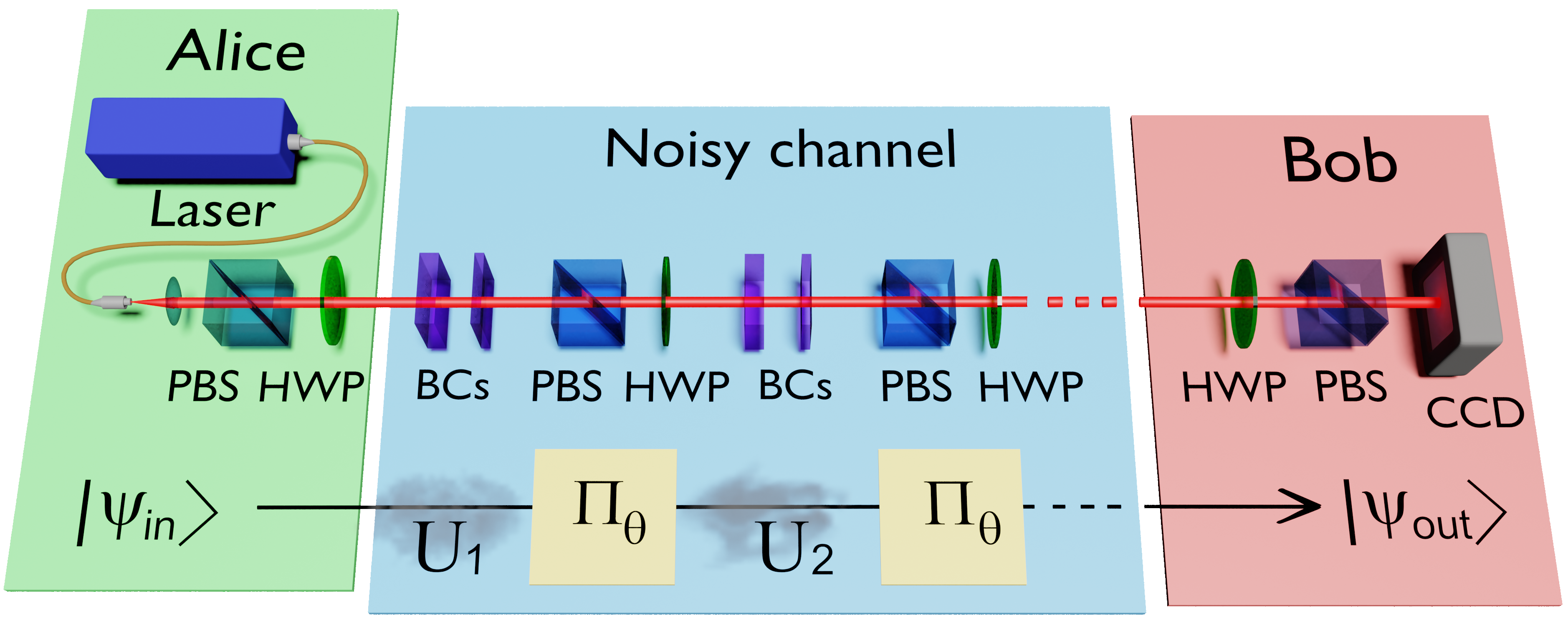}
	\caption{Experimental setup and corresponding quantum operations. \blue{For the sake of simplicity, only two of the six noise events occurring in the quantum channel of our experimental realization are shown (see main text for more details).}}
	\label{setup}
\end{figure}
We exploit a laser source at 700 nm attenuated down to the single-photon level, and arbitrarily choose as initial polarization state $\ket{\psi_{\theta}} = (\ket{H}+\ket{V})/\sqrt{2} \equiv \ket{+}$. We implement a series of $N=6$ decoherence-protection steps, each  composed of a birefringent crystal (BCs) pair positioned between two half-wave plates (HWPs) and followed by a polarizer (PBS), as shown in Fig. \ref{setup}. The first birefringent crystal (with optical axis cut at $45\un{^{\circ}}$ with respect to the photon propagation direction) shifts the horizontal polarization, as compared to the vertical one, in the transverse direction $x$. The shift depends on the crystal thickness and is associated to $G_k$.
\begin{figure*}[t!]
	\centering
	\includegraphics[width=0.9\textwidth]{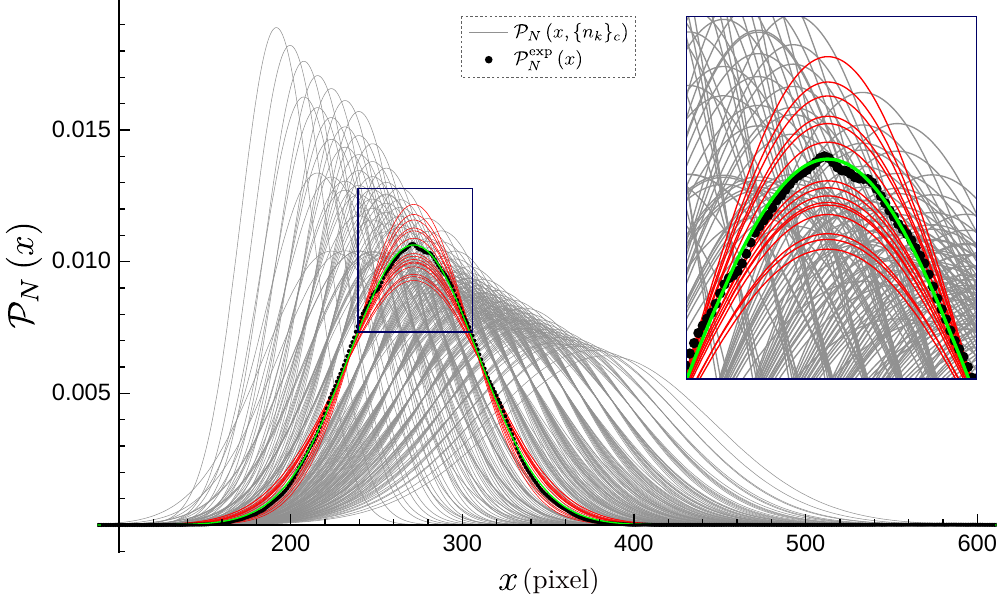}
	\caption{Experimental estimation of $\{n_k\}_R$. The measured spatial distribution $\mathcal{P}_N^{\mathrm{exp}}(x)$ is represented by the black dots, in very good agreement with the one given by the reconstructed set $\{n_k\}_R$ (green curve). The red curves correspond to the subset of probability configurations giving spatial distributions with same average position and different standard deviations, whereas the gray curves are the distributions due the remaining possible probability configurations. The inset in Fig.\,\ref{configuration} zooms in the distribution peak region, allowing to appreciate the correct discrimination performed within the subset of compatible noise configurations.}
	\label{configuration}
\end{figure*}
The second crystal (with optical axis at $90\un{^{\circ}}$) compensates for the temporal walk-off and phase mismatch between the $H$ and $V$ polarizations introduced by the previous crystal.
Then, the polarizer projects the photon onto its initial polarization state, thus inducing the QZE that protects the photon from decoherence.\\
At each step, we randomly choose among $D=5$ different crystal thicknesses corresponding to the set $\{G_k\} = (0, g, 2g, 3g, 4g)$, with $g$ being the minimum nonzero spatial displacement induced between the $H$ and $V$ polarization components.
The HWPs are needed in the initial setup calibration procedure for optimizing the temporal walk-off and phase compensation.\\
Finally, at the end of the $N$ decoherence-protection steps, the photons are detected by a CCD camera with $1024 \times 1024$ pixels of size $ (13 \times 13) \mu m$, providing the photons final spatial distribution $ f_{\rm out}(x)$.\\
In principle, the setup in Fig.\,\ref{setup} allows estimating any set $\{n_k\}$ by means of Eq.\,\eqref{setN}.
However, the minimization in Eq.\,\eqref{setN} is very challenging, since it requires an extremely precise measurement of the whole probability distribution profile, which is especially hard to achieve for the distribution tails.
Thus, in order to have a robust and reliable sensing procedure we compare, instead of the entire distributions, only some of their moments:
\begin{equation}
	\begin{split}
		E_{c}(x^i) &= \int \ud x \, x^i \mathcal{P}_N(x,\{n_k\}_c) \\
		E_{\mathrm{exp}}(x^i) &= \int \ud x \, x^i \mathcal{P}_N^{\mathrm{exp}}(x),
	\end{split}
\label{Ex}
\end{equation}
\blue{with $E_c (x^i)$ and $E_{exp}(x^i)$ being, respectively, the $i$-th moment of the theoretical distribution due to the set $\{n_k\}_c$ and the one extracted from the experimental spatial distribution $\mathcal{P}_N^{\mathrm{exp}}(x)$.}
In our case, we demonstrate that, thanks to our optimization approach, the first two moments suffice for faithfully extracting the $\{n_k\}_R$ set.\\
One can estimate the total amount of decoherence in the quantum channel by measuring the average position of $\mathcal{P}^{\mathrm{exp}}_N(x)$ both with and without exploiting the QZE, but this information is not enough to reconstruct the set $\{n_k\}_R$.
To do this, we start from a subset of the possible configurations that correspond to the estimated average position $E_{\mathrm{exp}}(x)$, and apply the minimization on the second-order moment difference $\Delta E_c(x^2) \equiv E_{\mathrm{exp}}(x^2) - E_c(x^2)$. As a result,
$\{n_k\}_R \simeq {\rm arg}\min_{c} \left(\Delta E_c(\Delta x^2)\right)^2$ with $\Delta x \equiv x - E_{\mathrm{exp}}(x)$.

\begin{figure}[t!]
	\centering
	\includegraphics[width=0.95\columnwidth]{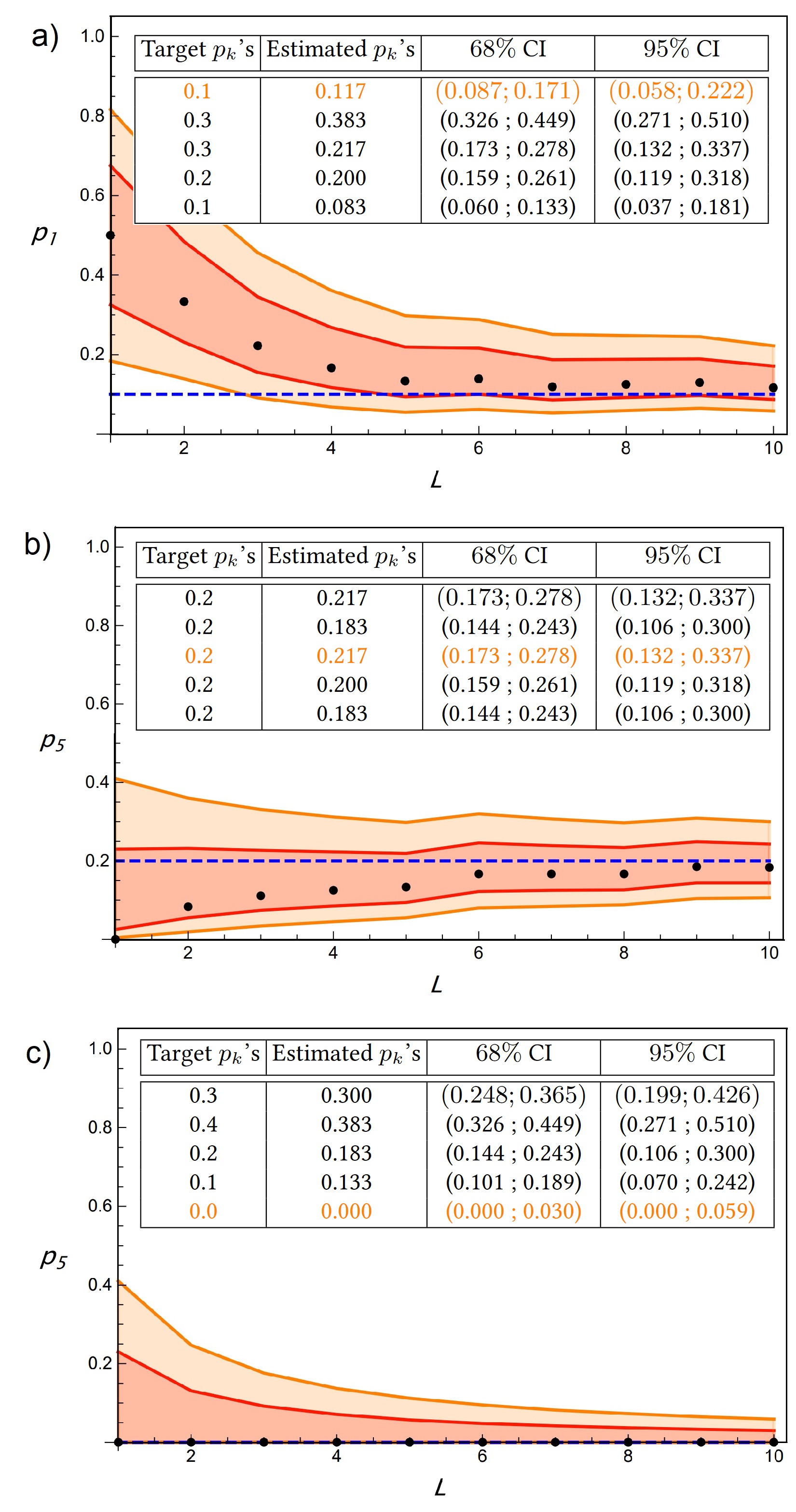}
	\caption{
	Considering three different noise probability distribution set respectively $\{p_k\}=(0.1,0.3,0.3,0.2,0.1)$ for panel (a), $\{p_k\}=(0.2,0.2,0.2,0.2,0.2)$ for panel (b), $\{p_k\}=(0.3,0.4,0.2,0.1,0.0)$ for panel (c).
	The tables report the reconstructed noise probability distribution $\{p_k\}_R$ for a chosen set $\{p_k\}$, together with the $68\%$ and $95\%$ confidence intervals (CIs) on the estimated values after six steps. \blue{As an example, the graphs show, for each set, the behaviour of a particular $p_k$ estimation (indicated in orange in the table) versus the number $L$ of procedure realizations. The blue dashed lines represent the theoretical $p_k$ value}, while the black dots are the estimated values. The red and orange areas correspond, respectively, to the $68\%$ and $95\%$ CI on the reconstructed $p_k$'s.}
	\label{Img_p}
\end{figure}

{\it Results --- }	
In Fig.\,\ref{configuration}, we present our results for a particular realization of the quantum channel noise distribution, corresponding to the crystal set $\{n_k\} = (2,0,2,2,0)$.
The experimental spatial distribution (black dots) is in very good agreement with the theoretical behavior (green curve) expected for the set $(2,0,2,2,0)$, correctly estimated by our sensing procedure.
The survival probability of the original polarisation state after six decoherence steps is $50 \%$ without protection, but it reaches $58 \%$ when we perform the noise sensing technique protecting the state.
Hence, we can sense the noise affecting the quantum channel without losing our probes in the channel, because of their higher survival probability.\\
In order to test the robustness and versatility of our technique, we apply it to three different noise event distributions (see Fig.\,\ref{Img_p}).
Then, to demonstrate the reduction of the uncertainty on the estimated $p_k$ values for an increasing number of trials $L$, we collect data for $L = 10$ randomized crystal sets $\{n_k\}$ for each chosen $\{p_k\}$.
Specifically, for each of the chosen noise distributions, $n_k$ is randomly sampled with probability $p_k$ for each $k=1,...,D$.
Furthermore, we estimate a confidence interval (CI) for each extracted
$\{p_k\}_R$ by exploiting the Beta distribution, which considers two different coverage factors associated to the $68\%$ and $95\%$ CIs, respectively.
The results are presented in the tables reported as insets in Fig. \ref{Img_p}, that correspond to the experimentally investigated  $\{p_k\}$ sets.
%
The estimated probabilities converge to the theoretical ones, within the $95\%$ CI, even after only $L=10$ realizations.
In addition, in Fig \ref{Img_p} we show, for each case, the typical behavior of the $p_k$ estimation results as a function of the number $L$ of the procedure realizations.
In each plot, the theoretical target probability is represented by a blue dashed line, while the estimated values obtained for each $L$ are shown by the black dots.
The experimentally extracted $\{p_k\}$ sets are in good agreement with the theoretical expectations in all the three cases considered, with most of the expected $p_k$'s falling within the $68\%$ CI on the corresponding reconstructed value.

{\it Conclusions --- }	
\new{In this paper we have proposed and demonstrated, both theoretically and experimentally, a novel noise sensing technique for unraveling, i.e. estimating, the statistics of stochastic decoherence events affecting a noisy channel.
Besides protecting from decoherence the (known) initial state of the qubit, QZE is shown to allow extracting information about the statistics of the decoherence source affecting the channel upon detecting the photon only at the end of the channel. 
Our technique is enabled by attributing different functionalities to distinct degrees of freedom of the employed quantum particles (here, photons) allowing them to serve both as quantum carriers and quantum sensors.\\
The proposed sensing technique has been experimentally demonstrated on a quantum optical setup, and has provided us with accurate results in each tested sequence of external decoherence events.
Since the photons are detected at the end of their optical path, our sensing procedure does not allow for the detection and reconstruction of noise temporal correlations, which would require detection also at intermediate times along the channel.\\
Our results open the way to a new generation of noise diagnostic tools, monitoring of microscopic noise events and eventually correcting for them by feedback.
A key application is in quantum communication, where the diagnostic capability offered by our procedure can provide a powerful real-time identification tool of failures and disruptions in the quantum channel, so that a prompt intervention is promoted.
This may be beneficial for distributed quantum computing and future quantum internet, in which flying qubits connect several noisy intermediate-scale quantum (NISQ) nodes \cite{PreskNISQ}.}


\section*{Acknowledgments}

This work was financially supported by the European Union's Horizon 2020 research and innovation programme under FET-OPEN Grant Agreement No. 828946 (PATHOS).
This work was also funded by the project QuaFuPhy (call ``Trapezio'' of Fondazione San Paolo) and by the projects EMPIR 19NRM06 METISQ and 20IND05 QADeT.
These last two projects received funding by the EMPIR program cofinanced by the Participating States and from the European Union Horizon 2020 Research and Innovation Program.
GK acknowledges support from DFG (FOR 2724) and QUANTERA (PACE-IN).

\bibliography{quantum2.bib}

\end{document}